\documentclass[%
 reprint,
 amsmath,amssymb,
 aps,
]{revtex4-2}

\usepackage{graphicx}% Include figure files
\usepackage{dcolumn}% Align table columns on decimal point
\usepackage{array}
\usepackage{bm}% bold math
\begin{document}

\preprint{APS/123-QED}

\title{A Simple and Accurate Method for Computing Optimized Effective \\
Potentials for Exact Exchange Energy}% Force line breaks with \\
%\thanks{A footnote to the article title}%

\author{Hideaki Takahashi}
% \altaffiliation[Also at ]{Physics Department, XYZ University.}%Lines break automatically or can be forced with \\
%\author{Second Author}%
 \email{ hideaki.takahashi.c4@tohoku.ac.jp}
\affiliation{%
Department of Chemistry, Graduate School of Science,\\
 Tohoku University, Sendai, Miyagi 980-8578, Japan 
}%

%\collaboration{MUSO Collaboration}%\noaffiliation

%\author{Charlie Author}
% \homepage{http://www.Second.institution.edu/~Charlie.Author}
%\affiliation{
% Second institution and/or address\\
% This line break forced with \\
%}
%\affiliation{
% Third institution, the second for Charlie Author
%}%
%\author{Delta Author}
%\affiliation{%
 %Authors' institution and/or address\\
 %This line break forced with \textbackslash\textbackslash
%}

%\collaboration{CLEO Collaboration}%\noaffiliation

\date{\today}% It is always \today, today,
             %  but any date may be explicitly specified

\begin{abstract}
The inverse Kohn-Sham density-functional theory (inv-KS) for the electron density of the Hartree-Fock (HF) wave function was revisited within the context of the optimized effective potential (HF-OEP). First, it is proved that the exchange potential created by the inv-KS is equivalent to the potential obtained by the HF-OEP when the HF-OEP realizes the HF energy of the system under consideration. Next the real-space grid (RSG) implementations of the inv-KS and the HF-OEP are addressed. The total HF energies $E^\text{HF}[\{\varphi_i^\text{inv-KS}\}]$ for the wave functions $\varphi_i^\text{inv-KS}$ on the effective potentials optimized by the inv-KS are computed for a set of small molecules. It is found that the mean absolute deviation (MAD) of $E^\text{HF}[\{\varphi_i^\text{inv-KS}\}]$ from the HF energy is clearly smaller than the MAD of $E^\text{HF}[\{\varphi_i^\text{HF-OEP}\}]$, demonstrating that the inv-KS is advantageous in constructing the detailed structure of the exchange potential $\upsilon_x$ as compared with the HF-OEP. The inv-KS method is also applied to an ortho-benzyne radical known as a strongly correlated polyatomic molecule. It is revealed that the spin populations on the atomic sites computed by the UHF calculation can be faithfully reproduced by the wave functions on the inv-KS potential.       
  
%\begin{description}
%\item[Usage]
%Secondary publications and information retrieval purposes.
%\item[Structure]
%You may use the \texttt{description} environment to structure your abstract;
%use the optional argument of the \verb+\item+ command to give the category of each item. 
%\end{description}
\end{abstract}

%\keywords{Suggested keywords}%Use showkeys class option if keyword
                              %display desired
\maketitle

%\tableofcontents

% \section{\label{sec:level1}The static correlation error in density functional theory}
\section{\label{sec:Introduction}Introduction}
The Kohn-Sham density-functional theory (KS-DFT)\cite{rf:hohenberg1964pr, rf:kohn1965pr} offers an efficient computational framework to study the electronic properties of materials and molecules. The KS-DFT has been established as a reliable and practical tool that can be applied to variety of systems.\cite{rf:parr_yang_eng,Martin2004} The success of KS-DFT is mainly due to the developments of the sophisticated exchange-correlation functionals $E_{xc}[n]$ given as explicit functionals of the electron density $n$. The simplest form of the exchange functional $E_{x}[n]$ was first developed on the basis of the local-density approximation (LDA)\cite{rf:kohn1965pr,rf:parr_yang_eng,Martin2004} using the homogeneous electron gas (HEG)\cite{rf:slater1951pr} as a model system.  Subsequent improvements on $E_{xc}[n]$ were made by adding terms with the density gradient $|\nabla n |$\cite{rf:becke1988pra, rf:perdew1996prl} and the second derivative $\nabla^2 n$\cite{rf:becke1989pra, rf:perdew1999prl} to consider the inhomogeneities of the real densities. Another major correction to $E_{xc}[n]$ was given by including a portion of the Hartree-Fock (HF) exchange energy to incorporate the difference in kinetic energy between the interacting real system and the corresponding non-interacting system.\cite{rf:becke1993Ajcp, rf:becke1993Bjcp} 

Since the set of wave functions $\{\varphi_i\}$ of the non-interacting reference system identified by the Kohn-Sham equation can be regarded as a functional of the ground state density, an orbital functional is an implicit density functional. Therefore, the HF energy functional $E^\text{HF}[\{\varphi_i\}]$ in terms of the KS wave functions can be considered as an implicit but exact functional within the framework of KS-DFT. The central issue in the orbital dependent functional for the HF exchange is how to optimize the \textit{local} exchange potential $\upsilon_x(\bm{r})$ that minimizes $E^\text{HF}[\{\varphi_i\}]$. The procedure have to be coupled with the optimization of the set of wave functions on the potential $\upsilon_x(\bm{r})$. The method is referred to as the Hartree-Fock optimized effective potential (HF-OEP),\cite{rf:talman1976pra,Gorling1999prl,Ivanov1999prl,Colle2001JPhysB,Sala2001jcp,rf:hirata2001jcp,rf:yang2002prl,kummel2003prl} which provides an implicit exact exchange functional $E_x$ in the KS-DFT. The exchange potential yielded by the HF-OEP can be directly compared with those given by approximate explicit $E_x$ functionals, where the HF-OEP potential serves as a reference for improving the functionals. The success of OEP not only justifies the approach of KS-DFT, but also proves the existence of an effective density functional. \\
\indent Apart from the OEP, the approach called inverse KS-DFT\cite{rf:leeuwen1994pra,Wu2003jcp,Kadantsev2004pra,Kanungo2019nat_com,Shi2021jpclett} offers a simple route to find the effective potential that uniquely corresponds to a given electron density without resorting to calculations with complex wave functions. The basic algorithm for the inverse KS-DFT is based on the variation principle proved by Foulkes and Haydock\cite{rf:foulkes1989prb} for a functional describing the kinetic energy of a non-interacting system under a constraint.  
At first, the inverse KS-DFT was utilized to construct a reference exchange-correlation potential for a given accurate electron density.\cite{rf:leeuwen1994pra,Wu2003jcp, Kadantsev2004pra,Kanungo2019nat_com,Shi2021jpclett} After the work by Wu and Yang,\cite{Wu2003jcp} the inverse KS-DFT also gained much concern within the context of the OEP.\cite{Shi2021jpclett}      
In the following, we make brief reviews of the HF-OEP and of the inverse KS-DFT for later discussions.    
\paragraph{HF-OEP}
The formulation of the OEP equation for the HF exchange was first given by Sharp and Horton\cite{rf:sharp1953pr} independent of the context of KS-DFT. Talman and Shadwick\cite{rf:talman1976pra} first solved the equation for atoms using the numerical grids. Krieger, Li, and Iafrate provided an approximation to the HF-OEP and applied it to a series of atoms showing excellent results.\cite{rf:krieger1992pra} G\"{o}rling and Levy formulated a basis set representation of the HF-OEP equation.\cite{Gorling1994pra} A lot of works have been devoted to develop numerical methods to solve the equation.\cite{Gorling1999prl,Ivanov1999prl,Colle2001JPhysB,Sala2001jcp,rf:hirata2001jcp,rf:yang2002prl,kummel2003prl} The solution of OEP for the HF energy requires the stationary condition of the HF energy $E^\text{HF}$ with respect to the variation of the objective effective potential $\upsilon_\text{eff}(\bm{r})$ for each spin,
\begin{equation}
\frac{\delta E^\text{HF}[\{\varphi_i\}]}{\delta \upsilon_\text{eff}(\bm{r})} = 0
\label{eq:dEHF_dveff}
\end{equation}
which leads to the following equation based on the 2nd-order perturbation theory
\begin{equation}
\frac{\delta E^{\text{HF}}\left[\left\{ \varphi_{i}\right\} \right]}{\delta\upsilon_{\text{eff}}\left(\bm{r}\right)}=\sum_{i}^{occ}\sum_{a}^{vir}\left\langle \varphi_{i}\right|\hat{H}_{\text{eff}}\left|\varphi_{a}\right\rangle \frac{\varphi_{a}^{*}\left(\bm{r}\right)\varphi_{i}\left(\bm{r}\right)}{\epsilon_{i}-\epsilon_{a}}=0
\end{equation}
where the operator $\hat{H}_{\text{eff}}$ is defined by 
\begin{equation}
\frac{\delta E^{\text{HF}}\left[\left\{ \varphi_{i}\right\} \right]}{\delta\varphi_{i}\left(\bm{r}\right)}=\hat{H}_{\text{eff}}\varphi_{i}^{*}\left(\bm{r}\right)
\end{equation}
and the wave functions $\{\varphi_{i}\}$ are the eigenfunctions of the KS equation with the local multiplicative potential $\upsilon_\text{eff}(\bm{r})$. 
In solving Eq. (\ref{eq:dEHF_dveff}) Yang and Wu(YW)\cite{rf:yang2002prl} proposed to introduce a fixed reference potential $\upsilon_0(\bm{r})$ to represent the long-range property of the exchange potential $\upsilon_x(\bm{r})$ in $\upsilon_\text{eff}(\bm{r})$. Then, the short range part of the exchange potential $\upsilon_x$ is expanded by a linear combination of a set of finite basis functions. The OEP problem is, thus, reduced to the optimization of the set of coefficients for the basis functions. YW approach is advantageous since it does not require the inversion of the response matrix $\chi(\bm{r},\bm{r}^\prime) = \frac{\delta n(\bm{r})}{\delta \upsilon_\text{eff}(\bm{r}^\prime)}$. Inverting $\chi(\bm{r},\bm{r}^\prime)$ suffers from the existence of the null eigenvector that corresponds to the fact that the effective potential includes an arbitrary additive constant.\cite{Ivanov1999prl} 

\paragraph{inverse KS-DFT}
In an approach developed by Wu and Yang,\cite{Wu2003jcp} the variation principle by Foulkes and Haydock\cite{rf:foulkes1989prb} based on the constrained search is exploited to generate the effective potential corresponding to an input electron density $n_\text{in}$. The constrained minimization is performed for the kinetic energy $\hat{T}$ of a non-interacting system thus
\begin{equation}
T_{s}\left[n_\text{in}\right]=\min_{\Psi_{\text{SD}}\rightarrow n_\text{in}}\left\langle \Psi_{\text{SD}}\right|\hat{T}\left|\Psi_{\text{SD}}\right\rangle 
\label{eq:Ts_func}
\end{equation}  
where $\Psi_\text{SD}$ is a Slater determinant and $n_\text{in}$ is supposed to be a $\upsilon$-representable density as an input to the functional $T_s$.  The constrained search Eq. (\ref{eq:Ts_func}) for the system of $N$ electrons can be written in another form by introducing a functional $\gamma [\Psi_\text{SD}]$,
\begin{align}
\gamma\left[\Psi_{\text{SD}}\right] & =2\sum_{i}^{N/2}\left\langle \varphi_{i}\right|-\frac{1}{2}\nabla^{2}\left|\varphi_{i}\right\rangle -\sum_{i}^{N/2}\epsilon_{i}\left(\left\langle \varphi_{i}\right|\left.\varphi_{i}\right\rangle -1\right)     \notag     \\
 & \;\;\;\;+\int d\bm{r}\lambda\left(\bm{r}\right)\left(n\left(\bm{r}\right)-n_{\text{in}}\left(\bm{r}\right)\right)
\label{eq:gamma}
\end{align}
where it is supposed that the wave function $\Psi_\text{SD}$ has a closed shell structure and is composed of a set of one-electron wave functions $\{\varphi_{i}\}\;(i=1,2, \cdots , N/2)$. 
$n_\text{in}$ is the objective density to be realized by the density $n\left(\bm{r}\right)=2\sum_{i}^{N/2}\varphi_{i}\left(\bm{r}\right)\varphi_{i}^{*}\left(\bm{r}\right)$ during the minimization of Eq. (\ref{eq:Ts_func}). The $\epsilon_i$ is the Lagrange's multiplier for the normalization condition of the wave function $\varphi_i$, and the function $\lambda(\bm{r})$ in the second line is that for the constraint of $n(\bm{r}) = n_\text{in}(\bm{r})$. The stationary condition of the functional $\gamma$ for the small variation of $\varphi_i$ is given by
\begin{equation}
\frac{\delta\gamma\left[\Psi_{\text{SD}}\right]}{\delta\varphi_{i}\left(\bm{r}\right)}=0\;\;\;\;\;\left(i=1,2,\cdots,N/2\right)
\label{eq:Stationary_cnd}
\end{equation}
which leads to Schr\"{o}dinger equations
\begin{equation}
\left(-\frac{1}{2}\nabla^{2}+\lambda\left(\bm{r}\right)\right)\varphi_{i}\left(\bm{r}\right)=\epsilon_{i}\varphi_{i}\left(\bm{r}\right)
\label{eq:Sch_eq}
\end{equation}
It is shown in the equation that $\lambda\left(\bm{r}\right)$ plays as a local potential to make the resultant density $n$ coincides with $n_\text{in}$. On the basis of this viewpoint, we switch to see $\gamma$ as a functional of the potential $\lambda\left(\bm{r}\right)$. Note that the potential $\lambda$ uniquely corresponds to the resultant density $n$ and the wave functions $\{\varphi_{i}\}$ through Eq. (\ref{eq:Sch_eq}). The potential $\lambda(\bm{r})$ that corresponds to $n_\text{in}$ can be obtained by the stationary condition
\begin{equation}
\frac{\delta\gamma\left[\lambda\right]}{\delta\lambda\left(\bm{r}\right)} = 0
\end{equation}    
This can be confirmed by the following formulation,
\begin{align}
\frac{\delta\gamma\left[\lambda\right]}{\delta\lambda\left(\bm{r}\right)} & =\left(2\sum_{i}^{N/2}\int d\bm{r}^{\prime}\frac{\delta\gamma\left[\lambda\right]}{\delta\varphi_{i}\left(\bm{r}^{\prime}\right)}\frac{\delta\varphi_{i}\left(\bm{r}^{\prime}\right)}{\delta\lambda\left(\bm{r}\right)}+c.c.\right)  \notag   \\
 & \;\;\;\;\;\;\;\;\;\;+n\left(\bm{r}\right)-n_{\text{in}}\left(\bm{r}\right)   \notag  \\
 & =n\left(\bm{r}\right)-n_{\text{in}}\left(\bm{r}\right)=0
\label{eq:FH_var} 
\end{align}
In deriving the second equality, the relations of Eq. (\ref{eq:Stationary_cnd}) are exploited. Thus it is revealed that the stationary condition is fulfilled when $n=n_\text{in}$ is achieved. Equation (\ref{eq:FH_var}) also indicates that $n\left(\bm{r}\right)-n_{\text{in}}\left(\bm{r}\right)$ constitutes the gradient vector to minimize $\gamma$. Such an approach to optimize the effective potential from a given electron density is often referred to as inverse Kohn-Sham (KS) DFT.\cite{Kadantsev2004pra, Kanungo2019nat_com, Shi2021jpclett} The second derivative of $\gamma$ with respect to $\lambda$ can also be evaluated to yield a Hessian $\frac{\delta^{2}\gamma}{\delta\lambda\left(\bm{r}\right)\delta\lambda\left(\bm{r}^{\prime}\right)}$ which can be used to expedite the convergence.\cite{Wu2003jcp} However, the inversion of the Hessian also suffers from the existence of the eigenvectors with zero or effectively zero eigenvalues,\cite{Heaton-Burgess2007prl, Bulat2007jcp} which give rise to ill-posed behaviors in the potentials. \\
\\
\indent In the present work, we first prove that the HF-OEP method is equivalent in principle to the inverse KS-DFT for the HF electron density $n^\text{HF}$ provided that the HF-OEP method realizes the HF total energy $E^\text{HF}$. To the best of our knowledge, this fact has not yet been clearly stated nor proved although the relationship was discussed in a literature.\cite{Bulat2007jcp} We perform our inverse KS-DFT as well as the HF-OEP calculations for small molecules to make comparisons and to show the equivalence of these methods. A notable feature of our implementation is that the real-space grids\cite{rf:chelikowsky1994prb,rf:chelikowsky1994prl,takahashi2001jpca,Takahashi2020jcim} combined with nonlocal pseudopotentials\cite{rf:kleinman1982prl} are utilized to express the wave functions and also the effective potentials. Heretofore, the atomic orbitals were mostly employed as the potential basis set,\cite{Gorling1994pra, Ivanov1999prl, rf:hirata2001jcp, rf:yang2002prl, Wu2003jcp, kummel2003prl, Heaton-Burgess2007prl, Bulat2007jcp, Shi2021jpclett} though the finite-element basis constructed from piecewise polynomials was also utilized in the work of Ref. \cite{Kanungo2019nat_com}. %To our knowledge, present work will be the first to construct the optimized potentials on the uniform grids. 

The first part of the next section is devoted to discuss the equivalency of the HF-OEP and the inverse KS-DFT. Then the theoretical details of the implementation of the HF-OEP and inverse KS-DFT with the real-space grids. In the third section, the results of the inverse KS-DFT are compared with those of HF-OEP and other work. In the last section, we discuss the advantage of the inverse KS-DFT in providing the local exact exchange potentials $\upsilon_x^\text{HF}(\bm{r})$ and the corresponding wave functions. A perspective is also provided for extending the HF-OEP or inverse KS-DFT to the density-functional theory based on a new formalism.\cite{rf:Takahashi2018, Takahashi2020jpb}        

\section{\label{sec:Methodology}Theory and Methods}
\subsection{HF-OEP and inverse Kohn-Sham DFT}   \label{subsec:OEP_vs_invKS}
\begin{figure}[h]
\centering
\scalebox{0.47}[0.47] {\includegraphics[trim=165 120 150 110,clip]{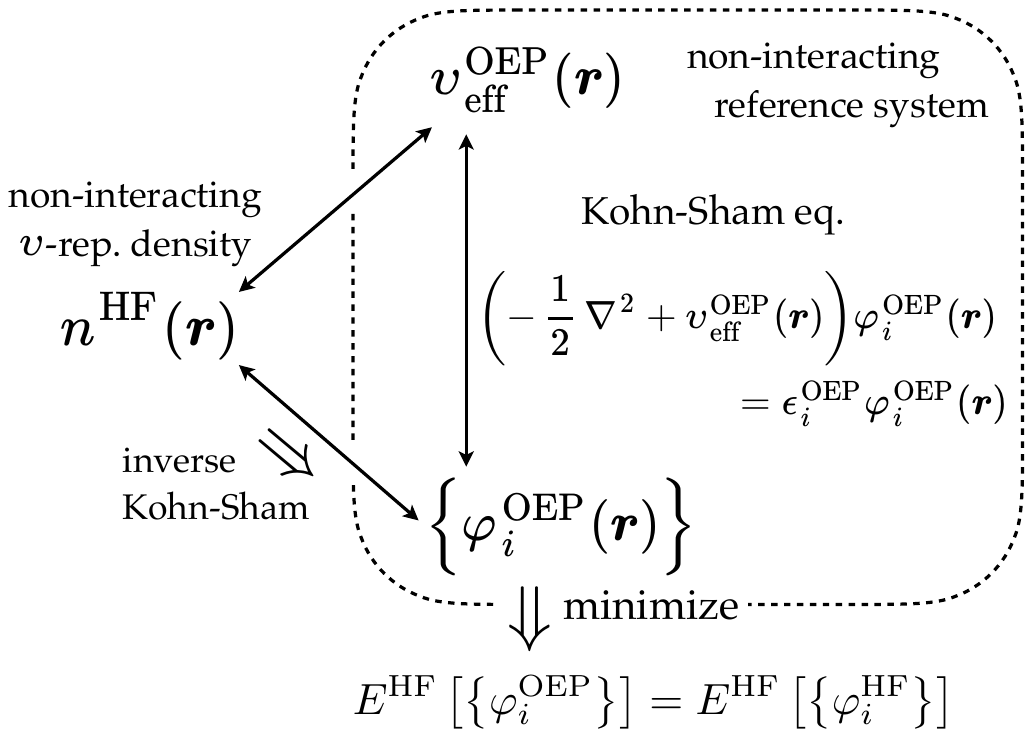}}            % Here is how to import EPS art
\caption{\label{fig:Triangle}The one-to-one correspondences among the optimized effective potential (OEP) $\upsilon_\text{eff}^\text{OEP}$ for the Hartree-Fock(HF) exchange, set of wave functions $\{\varphi_i^\text{OEP}\}$, and the ground state electron density $n^\text{HF}(\bm{r})$. The lines with arrows on both ends represent the one-to-one correspondences. }
\end{figure}

First we discuss the equivalence of HF-OEP and inverse Kohn-Sham DFT. To this end we consider a system with $N$ electrons in a closed shell structure. Suppose that a set of the one-electron wave functions $\{\varphi_i^\text{OEP}\}(i=1,2,\cdots,N/2)$ on the \textit{local} effective potential $\upsilon_\text{eff}^\text{OEP}(\bm{r})$ obtained through a HF-OEP calculation  minimizes the HF energy $E^\text{HF}[\{\varphi_i^\text{OEP}\}]$(see Fig. \ref{fig:Triangle}). The potential $\upsilon_\text{eff}^\text{OEP}(\bm{r})$ uniquely defines a non-interacting reference system with the wave functions $\{\varphi_i^\text{OEP}\}$ by the one-to-one correspondence.\cite{rf:hohenberg1964pr} Provided that $E^\text{HF}[\{\varphi_i^\text{OEP}\}]$ achieves the energy equal to the HF ground state energy $E^\text{HF}[\{\varphi_i^\text{HF}\}]$, the set of $\{\varphi_i^\text{OEP}\}$ is identical to the HF wave functions $\{\varphi_i^\text{HF}\}(i=1,2,\cdots,N/2)$ because of the variation principle. Then, $\{\varphi_i^\text{OEP}\}$ realizes the HF ground state density $n^\text{HF}(\bm{r})$. %On the other hand, provided that $n_0^\text{HF}(\bm{r})$ is non-interacting $\upsilon$-representable, the density $n_0^\text{HF}$ uniquely determines the set $\{\varphi_i\}(i=1,2,\cdots,N)$ of a non-interacting reference system. 
It directly indicates that the density $n^\text{HF}(\bm{r})$ is non-interacting $\upsilon$-representable.  
Therefore, the set of wave functions $\{\varphi_i\}$ obtained by the inverse Kohn-Sham method for the density $n^\text{HF}$ as the target is equivalent to the set $\{\varphi_i^\text{OEP}\}=\{\varphi_i^\text{HF}\}$. Thus the corresponding effective potential given by the inverse Kohn-Sham method is also equivalent to the potential given by the OEP method.       
%In the above discussion, the proof is based on the assumption that the HF-OEP can realize the HF energy $E^\text{HF}[\{\varphi_i^\text{HF}\}]$. Note, however, that the assumption is premised in the OEP calculation. Thus, the HF-OEP is equivalent to the inverse KS-DFT for the electron density $n^\text{HF}$ as an objective.       
%Figure1?
\subsection{Real-space Grid Implementation of HF-OEP}
The HF-OEP based on the YW approach\cite{rf:yang2002prl} is implemented on the real-space grids(RSG). In the RSG method,\cite{rf:chelikowsky1994prb,rf:chelikowsky1994prl,rf:takahashi2000cl, takahashi2001jpca} the KS equations for the valence electrons are written as
\begin{align}
\left(-\frac{1}{2}\nabla^{2}+\upsilon_{\text{H}}\left(\bm{r}\right)+\upsilon_{xc}\left(\bm{r}\right)+\upsilon_{\text{nuc}}^{\text{loc}}\left(\bm{r}\right)\right)\varphi_{i}\left(\bm{r}\right)\;\;\;\;      \notag    \\
+\left\langle \bm{r}\right|\hat{\upsilon}_{\text{nuc}}^{\text{nloc}}\left|\varphi_{i}\right\rangle =\epsilon_{i}\varphi_{i}\left(\bm{r}\right)
\label{eq:RSG_KS}
\end{align}
where $\upsilon_\text{H}$ and $\upsilon_{xc}$ are the Hartree and the exchange-correlation potentials, respectively. ${\upsilon}_{\text{nuc}}^{\text{loc}}$ and ${\upsilon}_{\text{nuc}}^{\text{nloc}}$ are, respectively, the local and the nonlocal parts of the pseudopotentials describing the interaction between the valence electrons and the nuclei with core electrons. All these potentials are represented on the uniform grids in a rectangular real-space cell. The Laplacian in the kinetic energy operator is expressed by the higher order finite-difference method.\cite{rf:chelikowsky1994prb,rf:chelikowsky1994prl} The base code for the RSG implementation is the \textquoteleft Vmol\textquoteright\; program\cite{rf:takahashi2000cl, takahashi2001jpca, rf:takahashi2001jcc, rf:takahashi2004jcp, takahashi2017jpcb, takahashi2019jpcb, Takahashi2020jcim} developed by the author and coworkers.     

Instead of the Fermi-Amaldi (FA) potential\cite{Zhao1994pra} employed as a fixed reference potential $\upsilon_0$ in the YW approach,\cite{rf:yang2002prl} in our RSG implementation, the potential $\upsilon_0$ is constructed by including the Slater's local exchange potential $\upsilon_{x}^{\text{Slater}}$,\cite{rf:slater1951pr}
\begin{equation}
\upsilon_{0}\left(\bm{r}\right)=\int d\bm{r}^{\prime}\frac{n^{\text{HF}}\left(\bm{r}^{\prime}\right)}{\left|\bm{r}-\bm{r}^{\prime}\right|}+\upsilon_{x}^{\text{Slater}}\left(\bm{r}\right)
\label{eq:ref_potential}
\end{equation} 
where $n^{\text{HF}}$ is the electron density of the HF ground state wave function of the system of interest. $\upsilon_{x}^{\text{Slater}}$ is expressed as
\begin{equation}
\upsilon_{x}^{\text{Slater}}\left(\bm{r}\right)=\sum_{i}^{occ}w_{i}\left(\bm{r}\right)\upsilon_{xi}^\text{loc}\left(\bm{r}\right)
\label{eq:Slater_loc_ex}
\end{equation}
$\upsilon_{xi}^\text{loc}$ is the local exchange potential specific to the orbital $\varphi_i^\text{HF}$ 
\begin{equation}
\upsilon_{xi}^\text{loc}\left(\bm{r}\right)=\frac{1}{\varphi_{i}^{\text{HF}}\left(\bm{r}\right)}\int d\bm{r}^{\prime}\frac{\sum_{j}^{N/2}\varphi_{j}^{\text{HF}}\left(\bm{r}\right)\varphi_{j}^{\text{HF}*}\left(\bm{r}^{\prime}\right)}{\left|\bm{r}-\bm{r}^{\prime}\right|}\varphi_{i}^{\text{HF}}\left(\bm{r}^{\prime}\right)
\end{equation}
where the wave functions $\varphi_i^\text{HF}$ are the solutions of the HF equation. 
The function $w_i(\bm{r})$ in Eq. (\ref{eq:Slater_loc_ex}) is the weight of the $i$th orbital at $\bm{r}$ and is given by
\begin{equation}
w_{i}\left(\bm{r}\right)=\frac{2\varphi_{i}^{\text{HF}}\left(\bm{r}\right)\varphi_{i}^{\text{HF*}}\left(\bm{r}\right)}{n^{\text{HF}}\left(\bm{r}\right)}
\end{equation} 
Note that the sum of the weight functions over the occupied orbitals becomes 1 regardless of $\bm{r}$ by definition. 
It was shown in our implementation that the reference potential $\upsilon_0$ given by Eq. (\ref{eq:ref_potential}) expedites the convergence of the SCF calculation in HF-OEP as compared with the FA potential. Of course, the long-range nature of the exchange potential can be ensured by the introduction of the reference potential.  

Now we introduce the grid function $g_p\left(\bm{r}\right)$ for the RSG implementation of the HF-OEP method. To do this, we consider the cubic fractional region $\Omega_p^h$ placed at $\bm{r}_p = (x_p, y_p, z_p)$ with the grid size $h$ in the real-space cell. Actually, $\Omega_p^h$ represents a spatial region that encloses a grid within the cell. Explicitly, the region $\Omega_p^h$ can be defined as
\begin{equation}
\Omega_{p}^{h} := \left\{ l_{p}-h/2\leq l\leq l_{p}+h/2\right\} \;\;\;\left(l=x,y,\text{and }z\right)
\end{equation}    
Then the grid function $g_p\left(\bm{r}\right)$ associated with the grid point $p$ is defined by
\begin{equation}
\begin{cases}
g_{p}\left(\bm{r}\right)=1 & \bm{r}\in\Omega_{p}^{h}      \\
g_{p}\left(\bm{r}\right)=0 & \bm{r}\notin\Omega_{p}^{h}
\end{cases}
\label{eq:grid_func}
\end{equation}
Using the grid functions $g_p$, the effective potential $\upsilon_\text{eff}^\text{Hx}$ due to the Hartree and the local exchange potential is expressed as
\begin{equation}
\upsilon_{\text{eff}}^{\text{Hx}}\left(\bm{r}\right)=\sum_{p}c_{p}g_{p}\left(\bm{r}\right)+\upsilon_{0}\left(\bm{r}\right)
\end{equation}    
where $c_p$ is the weight of the grid function $g_p$. In the following, the wave functions $\varphi_i^\text{OEP}$ are supposed to be optimized on the HF-OEP potential. For the sake of notational simplicity, however, the superscript OEP will be omitted hereafter.  
In parallel to the YW approach, the derivative of $E^\text{HF}[\{\varphi_i\}]$ with respect to $c_p$ is given by
\begin{align}
\frac{\partial E^{\text{HF}}\left[\left\{ \varphi_{i}\right\} \right]}{\partial c_{p}} & =-2\sum_{i}^{occ}\sum_{a}^{vir}\frac{\left\langle \varphi_{i}\left|g_{p}\right|\varphi_{a}\right\rangle }{\epsilon_{a}-\epsilon_{i}}    \notag    \\
 & \;\;\;\;\;\;\;\;\;\times\int d\bm{r}\frac{\delta E^{\text{HF}}\left[\left\{ \varphi_{i}\right\} \right]}{\delta\varphi_{i}\left(\bm{r}\right)}\varphi_{a}\left(\bm{r}\right)+c.c.
\label{eq:HFOEP_rsg}
\end{align}   
Within the RSG formalism, the integral $\left\langle \varphi_{i}\left|g_{p}\right|\varphi_{a}\right\rangle$ can be merely evaluated as $\varphi_{i}^{*}\left(\bm{r}_{p}\right)\varphi_{a}\left(\bm{r}_{p}\right)h^{3}$, where Eq. (\ref{eq:grid_func}) is used. Note that $\varphi_{i}\left(\bm{r}_{p}\right)$ is the value of the wave function $\varphi_{i}$ at the grid point $\bm{r}_{p}$. Thus the computational cost associated with the integration is quite minor in the RSG method. In the evaluation of $\int d\bm{r} \frac{\delta E^{\text{HF}}\left[\left\{ \varphi_{i}\right\} \right]}{\delta\varphi_{i}\left(\bm{r}\right)}\varphi_{a}\left(\bm{r}\right)$ in Eq. (\ref{eq:HFOEP_rsg}), the most time-consuming part is the application of the HF exchange potential to $\varphi_{a}\left(\bm{r}\right)$, that is,
\begin{equation}
\left\langle \bm{r}\left| \hat{\upsilon}_{xi}\right|\varphi_{a}\right\rangle =\varphi_{i}\left(\bm{r}\right)\int d\bm{r}^\prime \frac{\varphi_{i}^{*}\left(\bm{r}^{\prime}\right)\varphi_{a}\left(\bm{r}^{\prime}\right)}{\left|\bm{r}-\bm{r}^{\prime}\right|}
\end{equation}    
The operations must be computed for all the pairs $(i,a)$ between the occupied and the virtual orbitals. In our implementation, the calculation can be efficiently performed utilizing the Poisson solver based on the parallelized FFT.\cite{Takahashi2020jcim}  
The coefficients $c_p^k$ at the $k$th cycle are simply updated through 
\begin{equation}
c_p^{k+1} = c_p^k - \alpha \times \frac{\partial E^{\text{HF}}\left[\left\{ \varphi_{i}\right\} \right]}{\partial c_{p}^k}
\label{eq:update_oep}
\end{equation}
with $\alpha$ being a positive real number.

After the convergence of the HF-OEP, the local exchange potential $\upsilon_x^\text{OEP}(\bm{r})$ is obtained as
\begin{equation}
\upsilon_{x}^{\text{OEP}}(\bm{r})=\upsilon_{\text{eff}}^{\text{Hx}}\left(\bm{r}\right)-\int d\bm{r}^{\prime}\frac{n^{\text{OEP}}\left(\bm{r}^{\prime}\right)}{\left|\bm{r}-\bm{r}^{\prime}\right|}
\end{equation}
where $n^\text{OEP}$ is the electron density constructed from the orbitals $\varphi_i$ optimized on the converged HF-OEP potential. As discussed in subsection \ref{subsec:OEP_vs_invKS}, when the HF-OEP energy achieves $E^\text{HF}$, the electron density $n^\text{OEP}$ coincides with $n^\text{HF}$.     

\subsection{Real-space Grid Implementation of inverse KS-DFT}
The RSG implementation of the inverse KS-DFT is rather simple. Suppose that the HF electron density $n^\text{HF}\left(\bm{r}\right)$ is given from the outset. According to Eq. (\ref{eq:RSG_KS}), the KS equation at the $k$th optimization cycle of the inverse KS-DFT is written as
\begin{align}
\left(-\frac{1}{2}\nabla^{2}+\upsilon_{\text{eff}}^{k}\left[n^k\right]\left(\bm{r}\right)\right)\varphi_{i}\left(\bm{r}\right)  \notag  \\
\;\;\;\;+\left\langle \bm{r}\right|\hat{\upsilon}_{\text{nuc}}^{\text{nloc}}\left|\varphi_{i}\right\rangle  & =\epsilon_{i}\varphi_{i}\left(\bm{r}\right)
\end{align}
where $\upsilon_{\text{eff}}^{k}\left[n^k\right]\left(\bm{r}\right)$ can be decomposed into three terms
\begin{equation}
\upsilon_{\text{eff}}^{k}\left[n^k\right]\left(\bm{r}\right)=\upsilon_{x}^{k}\left(\bm{r}\right)+\upsilon_{\text{H}}\left[n^k\right]\left(\bm{r}\right)+\upsilon_{\text{nuc}}^{\text{loc}}\left(\bm{r}\right)
\end{equation}
By utilizing Eq. (\ref{eq:FH_var}) based on the Foulkes and Haydock variation principle, the local effective potential $\upsilon_\text{eff}^{k}\left(\bm{r}\right)$ can be updated for the next cycle by the following, 
\begin{equation}
\upsilon_\text{eff}^{k+1}\left(\bm{r}\right)=\upsilon_\text{eff}^{k}\left(\bm{r}\right)+\beta \left(n^k\left(\bm{r}\right)-n^{\text{HF}}\left(\bm{r}\right)\right)
\label{eq:update_revKS}
\end{equation}
with $\beta$ being a positive real number. On the potential $\upsilon_\text{eff}^{k+1}\left(\bm{r}\right)$ the new density $n^{k+1}$ is constructed. After the convergence of $\upsilon_\text{eff}\left(\bm{r}\right)$ and $n\left(\bm{r}\right)$, the exchange potential $\upsilon_x(\bm{r})$ can be obtained by
\begin{equation}
\upsilon_x\left(\bm{r}\right) = \upsilon_\text{eff}\left(\bm{r}\right) - \upsilon_\text{H}[n]\left(\bm{r}\right) - \upsilon_{\text{nuc}}^{\text{loc}}\left(\bm{r}\right)
\end{equation}
 The update of the potential using Eq. (\ref{eq:update_revKS}) is slow in general especially in the outer region of a molecule of interest because the electron density decays exponentially for the distance from the molecule. Fortunately, the problem can be efficiently alleviated by employing the Slater's local exchange potential $\upsilon_x^{\text {Slater}}(\bm{r})$ in Eq. (\ref{eq:Slater_loc_ex}) as an initial guess $\upsilon_x^{0}(\bm{r})$ of the potential realizing the proper long-range nature.  

To examine the equivalence of the HF-OEP and the inverse KS-DFT as discussed in subsection \ref{subsec:OEP_vs_invKS}, it is useful to monitor the energy $E^\text{HF}[\{\varphi_i^\text{inv-KS}\}]$ during the optimization cycle of the effective potential. It is shown in our calculations that the HF energy $E^\text{HF}[\{\varphi_i^\text{inv-KS}\}]$ in terms of the wave functions $\{\varphi_i^\text{inv-KS}\}$ optimized on the potential provided by the inverse KS-DFT decrease monotonically as the potential optimizations proceed. 

It is also worth noting that the RSG implementation of the inverse KS-DFT does not necessitate the nonlocal operations except for the nonlocal term of the pseudopotentials. Thus, the parallelization of the inverse KS-DFT on the real-space grid is quite straightforward, and it will be possible to achieve high parallel efficiency. Thus, we feel no need to expedite the optimization of the potential by using an inverted Hessian as proposed in Refs. \cite{Wu2003jcp, Bulat2007jcp}.     

\section{\label{sec:Comp_details}Computational Details}
\subsection{Real-space Grid Method}
Throughout the present KS-DFT calculations with the RSG method, the grid width $h$ is set at $h=0.22214$ a.u. For the atomic core regions, the double grid technique of Ref. \cite{rf:ono1999prl} is utilized to describe the rapid behaviors of the wave functions. The width of the dense grid is set at $h/7$. The pseudopotentials in Eq. (\ref{eq:RSG_KS}) are described with the separable form proposed by Kleinman and Bylander. \cite{rf:kleinman1982prl}  The kinetic energy operator is represented on the RSG with the 4th-order finite difference method.\cite{rf:chelikowsky1994prb,rf:chelikowsky1994prl,takahashi2001jpca,Takahashi2020jcim} The size of the cubic real-space cell is $L=26.657$ a.u., where the cell is uniformly discretized by 120 grids along each axis. The parallel computation of the KS-DFT with the RSG is performed using 16 CPUs through the  decomposition of the cell into $4\times 2\times 2$ domains.\cite{Takahashi2020jcim}    
\subsection{HF-OEP and inverse KS-DFT}
Both in the HF-OEP and the inverse KS-DFT calculations, the convergence threshold $\delta_\text{dns}$ in the electron density $n$ is set at $\delta_\text{dns} = 1.0\times10^{-5}$. Explicitly, the density is judged to be converged when a standard deviation satisfies the relation,  
\begin{equation}
\delta_\text{dns} \geq\left[\int d\bm{r}\left(n^{k+1}\left(\bm{r}\right)-n^{k}\left(\bm{r}\right)\right)^{2}\right]^{\frac{1}{2}}
\end{equation}
The integration is performed by the discrete sum over the grid points in the real-space cell. 
 For the HF-OEP calculations, threshold is also provided for the effective potential and is set at $\delta_\text{pot} = 1.0\times10^{-5}$.    
The acceleration parameters $\alpha$ and $\beta$ in Eqs. (\ref{eq:update_oep}) and (\ref{eq:update_revKS}) are set at $5.0$ and $0.8$, respectively, in the present work.  

\section{Results and Discussions}
\subsection{Inverse Kohn-Sham DFT}
To examine the performance of the inverse KS-DFT(inv-KS) implemented on the real-space grids, we use the electron densities yielded by the LDA(local-density approximation) exchange functional\cite{rf:slater1951pr} and a GGA(generalized gradient approximation) exchange-correlation functional. Since these functionals provide the local potentials $\upsilon_{xc}$ on the real space, the potentials serve as references for the inv-KS calculations. We consider the densities of CH$_2$ molecule with the singlet spin state. The valence orbitals constructed in a full electrons calculation using an LCAO basis set are employed as the initial guess for the wave functions. The LDA exchange potential is used as the initial guess for the potentials $\upsilon_{xc}$ for these test calculations instead of the Slater's local potential (Eq. (\ref{eq:Slater_loc_ex})) since the LDA and GGA potentials are known to be short range. The BLYP functional\cite{rf:becke1988pra,rf:lee1988prb} is utilized as a GGA functional in the present calculation. 

\begin{figure}[h]
\centering
\scalebox{0.33}[0.33] {\includegraphics[trim=3 10 15 0, clip]{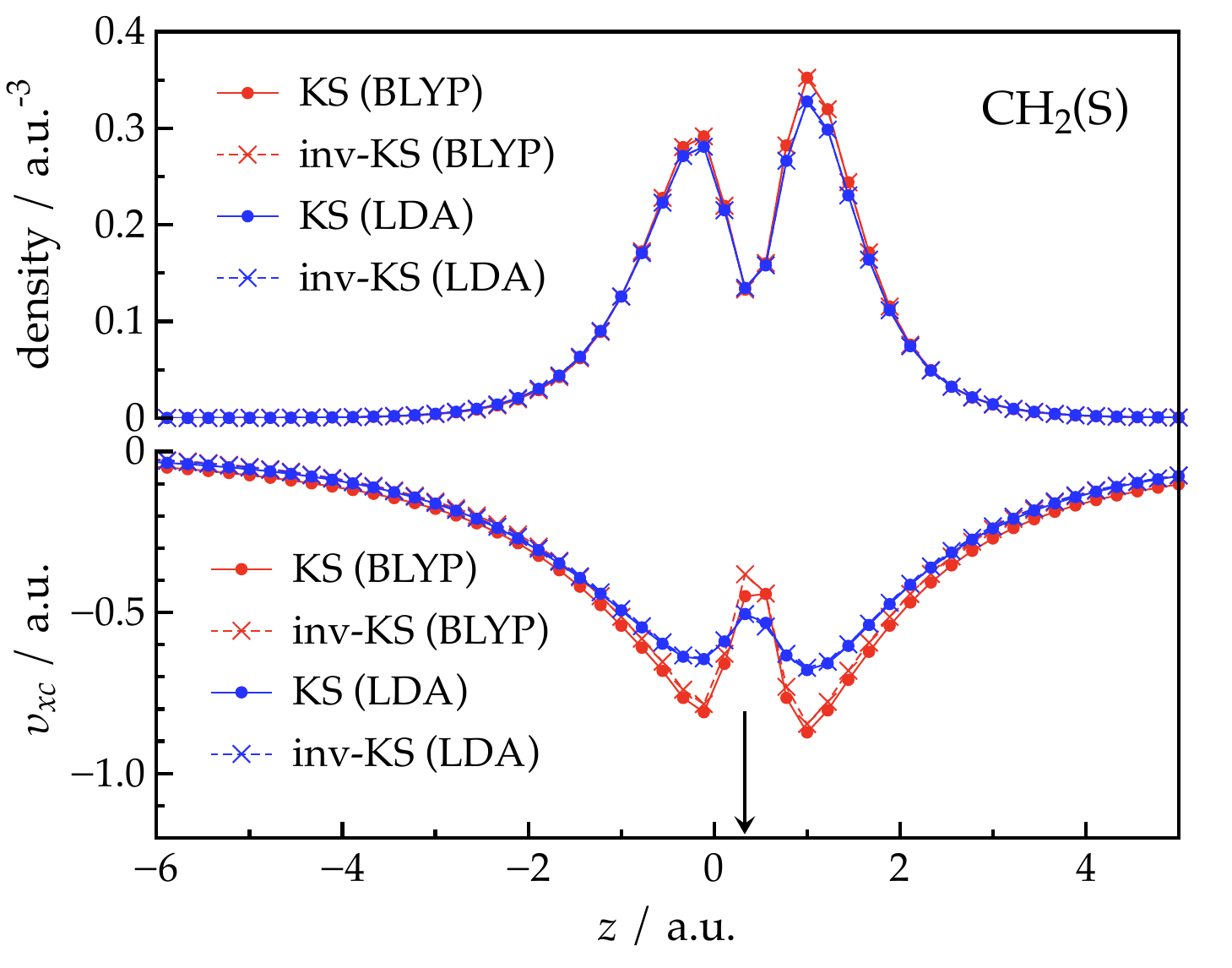}}            % Here is how to import EPS art
\caption{\label{fig:KS_vs_revKS_CH2} The density $n(\bm{r})$ (upper) of the valence electrons and the exchange-correlation potential $\upsilon_{xc}(\bm{r})$ (lower) on the molecular axis of the singlet ground state CH$_2$ molecule. The molecular axis is aligned parallel to the $z$ axis. The arrow indicates the position of C atom on the $z$ axis. The depressions in the densities and the potentials around the C atom are due to the fact that only valence electrons are considered in the calculations. The molecular geometry is that given in Ref. \cite{Ivanov1999prl}. }
\end{figure}

The results are shown in Fig. \ref{fig:KS_vs_revKS_CH2}, where the potentials $\upsilon_{xc}$ as well as the densities along the molecular axis are presented. It is shown that the exchange-correlation potential by BLYP functional has a more distinct structure as compared with the LDA exchange potential although the resultant electron densities do not differ significantly from each other. The densities provided by the inv-KS calculations using the real-space grids are shown to faithfully reproduce the LDA and the GGA densities. It is also demonstrated that the LDA exchange potential provided by KS-DFT can be realized by the inv-KS calculation. It is found, however, that the potential $\upsilon_{xc}$ obtained by inv-KS slightly differs from that given by KS-DFT. As shown in the figure, the potential $\upsilon_{xc}$ by the KS-DFT calculation with the BLYP functional has a long-range nature as compared with the LDA exchange exchange potential. Since the LDA potential is used as the initial guess for the inv-KS calculation for the BLYP density, the long-range nature cannot be realized. However, the difference between the total energies $E^\text{BLYP}_\text{tot}[n^\text{inv-KS}]$ and $E^\text{BLYP}_\text{tot}[n^\text{KS}]$ is found to be less than $10^{-2}$ mHartree. Thus, the reliability of the present implementation of the inverse KS-DFT on the real-space grids is demonstrated. 
%Figure2?

\subsection{Inverse Kohn-Sham DFT and HF Optimized Effective Potential}
As discussed in subsection \ref{subsec:OEP_vs_invKS}, the inv-KS provides the effective potential equivalent to the potential obtained by the HF-OEP method when the HF-OEP realizes the HF total energy $E^\text{HF}$. Here we examine the accuracy of the inv-KS method when it is applied to the electron densities given by the HF method. The set of molecules for the test is that provided in the work by Yang and Wu (YW).\cite{rf:yang2002prl} The results of the inv-KS are directly compared with those given by YW. To see the effects of the real-space grids used as the basis set to describe the effective potential and the wave functions, HF-OEP calculations with the real-space grids are also performed. The results are summarized in Table \ref{tab:OEP_vs_invKS}. The energy $E(\text{inv-KS})$ is computed as $E^\text{HF}[\{\varphi_i^\text{inv-KS}\}]$ where $\varphi_i^\text{inv-KS}$ are the wave functions on the converged effective potential in the inv-KS. It is shown in the table that the implementation of the HF-OEP in this work is slightly worse in accuracy than the results by YW although almost comparable. On the other hand, it is found that the inv-KS calculations give the better results than the HF-OEP calculations for every molecules tested in this study. It is apparent that the inv-KS is better in accuracy than the HF-OEP at least for the set of these 14 molecules. Actually, the mean absolute deviation (MAD) of the inv-KS is distinctly smaller than that of the HF-OEP. Furthermore, the deviation of the energy $E^\text{HF}\left[\{\varphi_i^\text{inv-KS}\}\right]$ from the HF energy for each molecule is found to be always smaller than that of the energy $E^\text{HF}\left[\{\varphi_i^\text{HF-OEP}\}\right]$ for the test set. This also holds when the inv-KS energies are compared with the HF-OEP energies of other works by Yang and Wu\cite{rf:yang2002prl} or by Ivanov et al.\cite{Ivanov1999prl} %The advantage of inv-KS can be attributed to the fact that the potential is simply updated using Eq. (\ref{eq:FH_var}). Since the response function $\chi(\bm{r},\bm{r}^\prime)$ nor its inverse is not needed in the inv-KS method, the references to the virtual orbitals are not necessary.     
\begin{table}[h!]
\caption{\label{tab:OEP_vs_invKS} The energy differences of $E$(OEP)$-E$(HF) and $E$(inv-KS)$-E$(HF) in the units of mHartree. OEP(YW): Yang and Wu\cite{rf:yang2002prl}, OEP(PW):  real-space grid implementation of OEP in the present work, inv-KS(PW): inverse KS-DFT in the present work. MAD: mean absolute deviation. The molecular geometries are those provided in Refs. \cite{Ivanov1999prl,rf:yang2002prl}}
\centering
%\begingroup
\renewcommand{\arraystretch}{1.2}
%\begin{tabular}{c c c c}
\begin{tabular}{p{6.0em}p{6em}p{6em}p{6.8em}}
\hline 
{molecule} & \centering{OEP(YW)}  & \centering{OEP(PW)} & \centering{inv-KS(PW)} \tabularnewline
{H$_2(S=1)$} & \centering{$0.00$} & \centering{$0.00$} & \centering{$0.00$}  \tabularnewline
{H$_2$O$(S=1)$} & \centering{$2.27$} & \centering{$2.37$} & \centering{$1.47$}  \tabularnewline
{HF$(S=1)$} & \centering{$1.95$} & \centering{$2.43$} & \centering{$1.50$}  \tabularnewline
{OH$(S=2)$} & \centering{$2.38$} & \centering{$2.69$} & \centering{$1.64$}  \tabularnewline
{N$_2(S=1)$} & \centering{$5.20$} & \centering{$5.03$} & \centering{$3.88$}  \tabularnewline
{O$_2(S=3)$} & \centering{$6.52$} & \centering{$7.93$} & \centering{$5.56$}  \tabularnewline
{F$_2(S=1)$} & \centering{$8.51$} & \centering{$10.3$} & \centering{$7.94$}  \tabularnewline
{CH$_2(S=1)$} & \centering{$2.90$} & \centering{$3.49$} & \centering{$1.81$}  \tabularnewline
{CH$_2(S=3)$} & \centering{$1.87$} & \centering{$1.45$} & \centering{$0.90$}  \tabularnewline
{NH$_2(S=2)$} & \centering{$2.55$} & \centering{$2.72$} & \centering{$1.61$}  \tabularnewline
{NH$(S=3)$} & \centering{$1.97$} & \centering{$2.35$} & \centering{$1.25$}  \tabularnewline
{CO$(S=1)$} & \centering{$5.12$} & \centering{$5.18$} & \centering{$3.84$}  \tabularnewline
{CN$^{-}(S=1)$} & \centering{$4.41$} & \centering{$4.69$} & \centering{$3.21$}  \tabularnewline
{OH$^{-}(S=1)$} & \centering{$2.14$} & \centering{$3.53$} & \centering{$1.46$}  \tabularnewline
\\ [-3mm]
\centering{MAD} & \centering{$3.41$} & \centering{$3.87$} & \centering{$2.58$}  \tabularnewline
\hline 
\end{tabular}
\end{table}      

\begin{figure}[h]
\centering
\scalebox{0.33}[0.33] {\includegraphics[trim=3 10 15 0, clip]{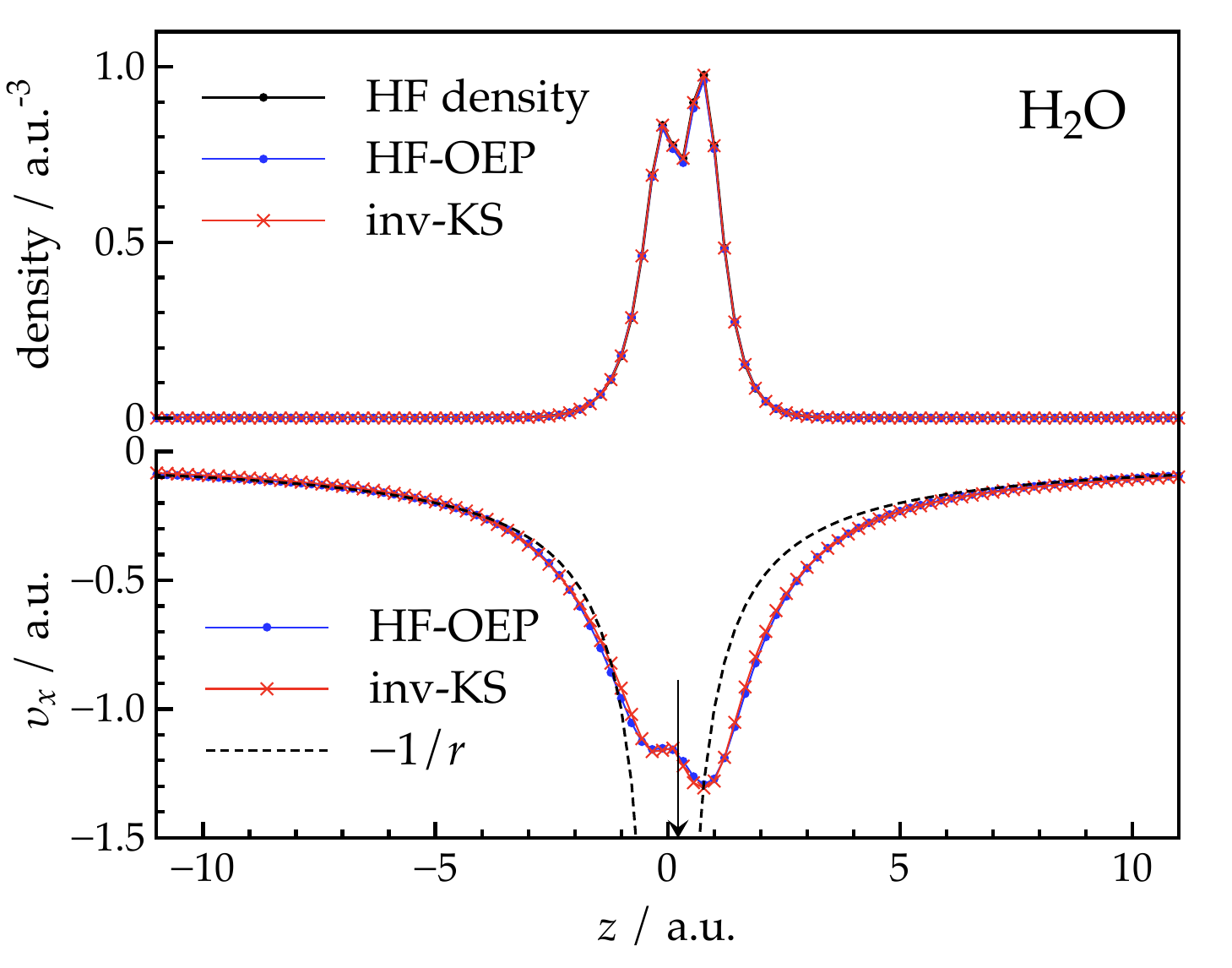}}            % Here is how to import EPS art
\caption{\label{fig:OEP_vs_revKS_H2O} The density $n(\bm{r})$ (upper) of the valence electrons and the exchange-correlation potential $\upsilon_{xc}(\bm{r})$ (lower) on the molecular axis of the singlet ground state H$_2$O molecule. The molecular axis is aligned parallel to the $z$ axis. The arrow indicates the position of O atom on the $z$ axis. The molecular geometry is that given in Ref. \cite{Ivanov1999prl}. $r$ is the distance from the origin of the $z$ axis. }
\end{figure}

We also make a comparison between the exchange potentials $\upsilon_x$ obtained by the HF-OEP and the inv-KS calculations for an H$_2$O molecule. In Fig. \ref{fig:OEP_vs_revKS_H2O}, the electron densities $n(\bm{r})$ and the exchange potentials $\upsilon_x(\bm{r})$ along the molecular axis are presented. For the densities, it is shown in the graphs that the HF-OEP and the inv-KS methods are both successful in reproducing the target HF electron density. The exchange potential by HF-OEP differs only slightly from that by inv-KS around the oxygen atom. It is notable that these two potentials both show the correct asymptotic behavior of $-1/r$. This is the direct consequence of the fact that the Slater's local exchange potential in Eq. (\ref{eq:Slater_loc_ex}) is incorporated in the reference potentials for these methods.       
%Figure3

\begin{figure}[h]
\centering
\scalebox{0.49}[0.49] {\includegraphics[trim=220 10 80 5,clip]{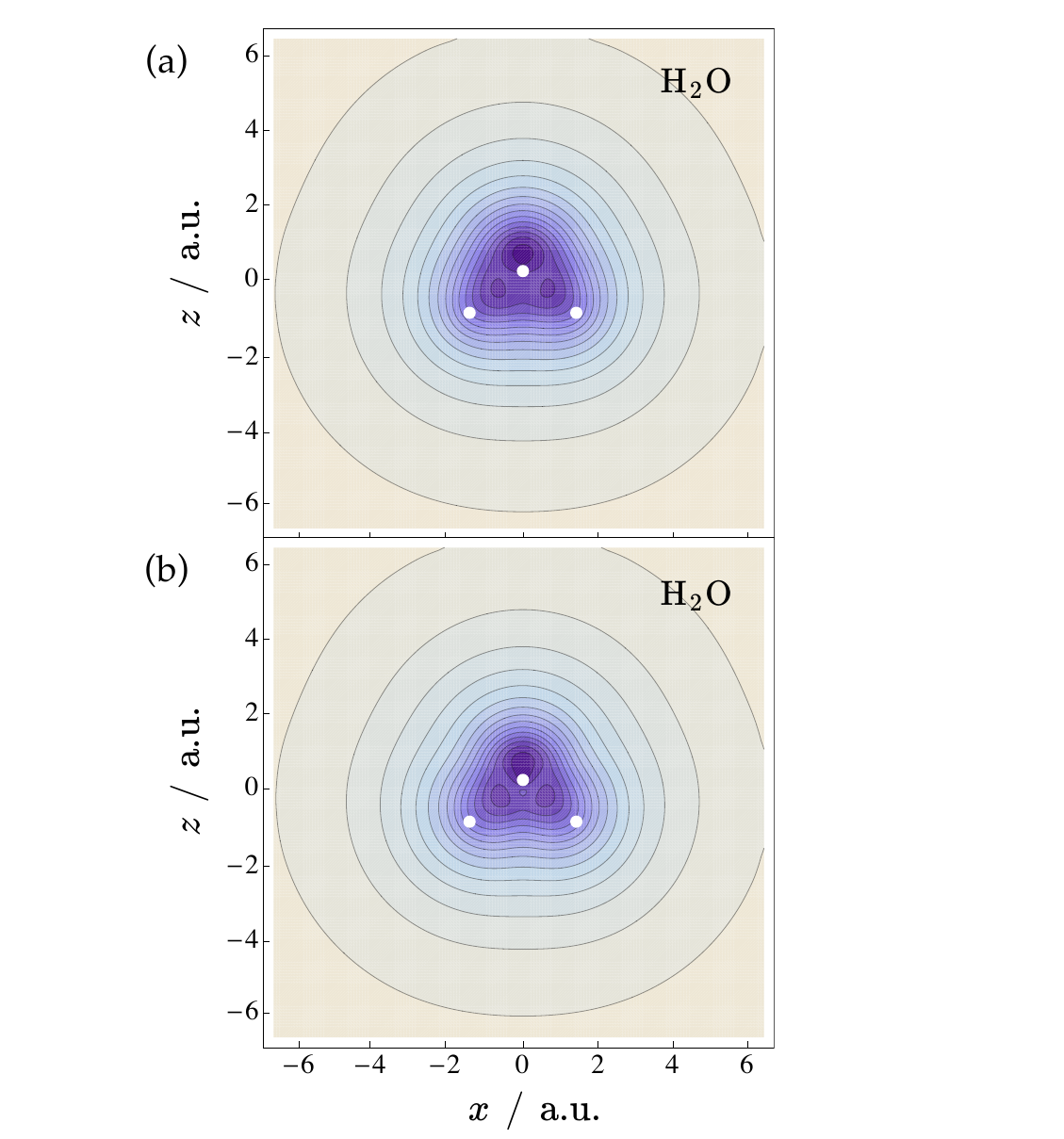}}            % Here is how to import EPS art
\caption{\label{fig:OEP_vs_invKS_contour} Contour plots of the exchange potentials $\upsilon_x(\bm{r})$ optimized by (a) the HF-OEP and (b) the inverse KS methods on the H$_2$O molecular plane placed on the $xz$ plane. The darker region is lower in the energy. The energy range of the contour lines is from -1.28 a.u. to -0.16 a.u. The interval of the adjacent contour lines is 0.08 a.u. The white dots indicate the positions of the oxygen and hydrogen atoms in the water molecule. The molecular geometry is that given in Ref. \cite{Ivanov1999prl}. }
\end{figure}

To make detailed comparisons between the potentials obtained by the HF-OEP and the inv-KS methods, we provide in Figs. \ref{fig:OEP_vs_invKS_contour}(a) and \ref{fig:OEP_vs_invKS_contour}(b) the contour plots of the exchange potentials on the molecular plane, respectively. The overall coincidence between the potentials is notable. However, it is observed in the figure that the detailed structure of the potential given by the HF-OEP is somewhat different from that by the inv-KS, which results in the difference in the energy errors of $2.37$ and $1.47$ mHartree shown in Table \ref{tab:OEP_vs_invKS}. It should also be noted that the error $2.37$ mHartree (=$E$(OEP)$-E$(HF)) of the present work is comparable to the error ($2.27$ mHartree) of the work by Yang and Wu(YW).\cite{rf:yang2002prl}  And the result provided by YW is also comparable to that by another OEP calculation of Ivanov et al\cite{Ivanov1999prl} where the error was evaluated as $2.30$ mHartree. The disadvantage of the HF-OEP can be attributed to the use of the response function $\chi(\bm{r},\bm{r}^\prime)$ in the construction of the exchange potential, which necessitates the references to the virtual orbitals within a truncated subspace. On the other hand, in the inv-KS approach, the exchange potential can be directly updated by the increment of the fraction of the density difference $n-n^\text{HF}$ according to Eq. (\ref{eq:FH_var}). This will be advantageous in constructing the detailed structures of the effective potentials.           
%Figure4
  
\subsection{Spin-polarized Polyatomic System}
Heretofore, the number of calculations of the HF-OEP or the inverse KS-DFT applied to realistic polyatomic systems is quite small. To the best of our knowledge, the work by Kanungo et al\cite{Kanungo2019nat_com} was the first that applied the inv-KS method based on the finite-element basis to weakly and strongly correlated molecular systems that include up to 58 electrons. Here we apply our inv-KS method based on the real-space grid basis to the ortho-benzyne radical (C$_6$H$_4$) known as a strongly correlated system. The diradical character arising from the static correlation on the adjacent two carbon atoms will be described utilizing the symmetry-broken (spin-unrestricted) HF wave functions in the present approach. The difference $E^\text{HF}(\text{inv-KS})-E^\text{HF}(\text{UHF})$ in the total energy between the inv-KS and the HF methods is obtained as $13.5$ mHartree. Thus the error of the wave functions on the effective local potential is found to be rather large as compared with those of the molecules listed in Table \ref{tab:OEP_vs_invKS}. However, the magnitude of the error can be mainly attributed to the system size since the error is additive with respect to the number of valence electrons included in the system. According to this consideration, it can be readily confirmed that the error of the inv-KS calculation for C$_6$H$_4$ is comparable to that for the O$_2$ molecule shown in Table \ref{tab:OEP_vs_invKS}. The wave function on the exchange potential optimized by the inv-KS yields the dipole moment $d$ of the system as $d= 1.321$ Debye, while it is computed as $d=1.331$ Debye in the UHF calculation.  
%Figure5? 
\begin{figure}[h]
\centering
\scalebox{0.51}[0.51] {\includegraphics[trim=220 10 80 5,clip]{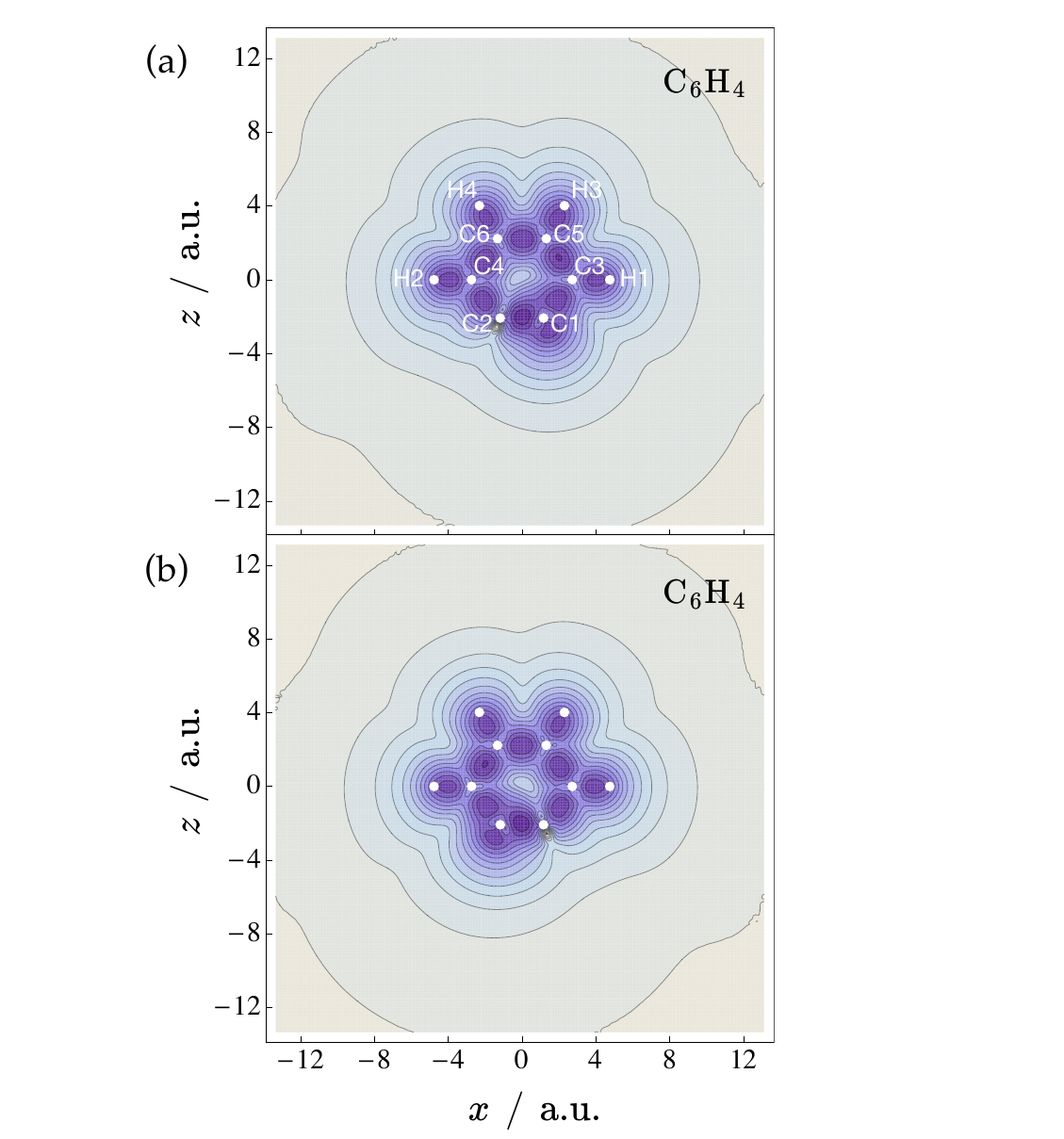}}            % Here is how to import EPS art
\caption{\label{fig:invKS_c6h4} Contour plots of the exchange potentials for (a) the $\alpha$ and (b) the $\beta$ spin electrons optimized by the inverse KS methods on the molecular plane of C$_6$H$_4$ placed on the $xz$ plane. The darker region is lower in the energy. The energy range of the contour lines is from -1.04 a.u. to -0.08 a.u. The interval of the adjacent contour lines is 0.08 a.u. The white dots indicate the positions of the carbon and hydrogen atoms in the C$_6$H$_4$ molecule. The molecular geometry is that provided in the Supplementary material of Ref. \cite{Kanungo2019nat_com}. Numberings on the atoms are indexes for references in the main text. }
\end{figure} 

In Figs. \ref{fig:invKS_c6h4}(a) and \ref{fig:invKS_c6h4}(b), the exchange potentials $\upsilon_{x\sigma}(\bm{r})$ for spins $\sigma = \alpha \text{ and }\beta$ in C$_6$H$_4$ are presented, respectively. Interestingly, it is observed in the figure that the potentials $\upsilon_{x\sigma}(\bm{r})$ differ from each other significantly only at the two major radical sites C1 and C2. We evaluate the spin populations of the wave functions on the effective potential optimized by the inv-KS approach. Explicitly, the spin populations assigned to the atomic sites are quantified by the fuzzy cell method\cite{rf:becke1988Ajcp} and the values are compared with those obtained by the UHF calculation. The results are summarized in Table \ref{tab:spin_pop_c6h4}. It is seen in the table that an alternation of the $\alpha$ and the $\beta$ spin densities emerges along the six-membered ring composed of the carbon atoms, where C1 and C2 atoms have the largest spin populations($\pm1.061$) with anti-parallel directions. It should be noted that the spin densities obtained through the inv-KS method show good agreements with those given by the corresponding UHF calculation. This indicates the accuracy and the reliability of the exchange potential $\upsilon_{x\sigma}$ obtained by the application of the present inv-KS method to a strongly correlated polyatomic system.  

\begin{table}[h!]
\caption{\label{tab:spin_pop_c6h4} Spin populations assigned to atomic sites computed by the inverse KS-DFT and the UHF methods. Refer to the atom indexes presented in Fig. \ref{fig:invKS_c6h4}(a). In the fuzzy cell method to assign the spin populations to atoms, the atomic radii proposed in Ref. \cite{Slater1964jcp} are used. }
\centering
%\begingroup
\renewcommand{\arraystretch}{1.3}
\begin{tabular}{c r r}
%\begin{tabular}{c D{.}{.}{5} D{.}{.}{5}}
%\begin{tabular}{m{25mm}b{25mm}b{25mm}}
\hline 
{\;\;\;\;\;\;\;\;atomic site\;\;\;\;\;\;\;\;} & {\;\;\;\;\;\;\;inv-KS\;\;\;\;\;\;\;\;\;\;\;\;}  & {\;\;\;\;UHF\;\;\;\;\;\;\;}  \\
{C$1$} & {$1.061$\;\;\;\;\;\;\;\;\;\;\;\;\;} & {$1.062$\;\;\;\;\;\;\;}   \\
{C$2$} & {$-1.061$\;\;\;\;\;\;\;\;\;\;\;\;\;} & {$-1.062$\;\;\;\;\;\;\;}   \\
{C$3$} & {$-0.424$\;\;\;\;\;\;\;\;\;\;\;\;\;} & {$-0.426$\;\;\;\;\;\;\;}   \\
{C$4$} & {$0.424$\;\;\;\;\;\;\;\;\;\;\;\;\;} & {$0.426$\;\;\;\;\;\;\;}   \\
{C$5$} & {$0.418$\;\;\;\;\;\;\;\;\;\;\;\;\;} & {$0.418$\;\;\;\;\;\;\;}   \\
{C$6$} & {$-0.418$\;\;\;\;\;\;\;\;\;\;\;\;\;} & {$-0.418$\;\;\;\;\;\;\;}   \\
{H$1$} & {$0.018$\;\;\;\;\;\;\;\;\;\;\;\;\;} &  {$0.016$\;\;\;\;\;\;\;}   \\
{H$2$} & {$-0.017$\;\;\;\;\;\;\;\;\;\;\;\;\;} & {$-0.016$\;\;\;\;\;\;\;}   \\
{H$3$} & {$-0.005$\;\;\;\;\;\;\;\;\;\;\;\;\;} & {$-0.004$\;\;\;\;\;\;\;}   \\
{H$4$} & {$0.005$\;\;\;\;\;\;\;\;\;\;\;\;\;} & {$0.004$\;\;\;\;\;\;\;}   \\
\hline 
\end{tabular}
\end{table}      

\section{conclusion}
In this work it was proved that the HF exchange potential obtained by the inverse KS-DFT(inv-KS) is equivalent to that obtained by the HF-OEP method provided that the HF-OEP realizes the HF ground state energy of a system under consideration. We implemented the HF-OEP and the inv-KS methods utilizing the real-space grids (RSGs) and the pseudopotentials for valence electrons. It was demonstrated through the test calculations for small molecules that the inv-KS method is superior in accuracy to the HF-OEP. %Actually, the deviation of the energy $E^\text{HF}\left[\{\varphi_i^\text{inv-KS}\}\right]$ from the HF energy is found to be always smaller than that of the energy $E^\text{HF}\left[\{\varphi_i^\text{HF-OEP}\}\right]$ for the test set. The mean absolute deviation (MAD) of the inv-KS is distinctly smaller than that of the HF-OEP. This also holds when the inv-KS energies are compared with the HF-OEP energies of other works by Yang and Wu\cite{rf:yang2002prl} or by Ivanov et al.\cite{Ivanov1999prl} 
The advantage of the inv-KS can be attributed to its method to update the effective potential, where the use of the response function is not needed by virtue of the Foulkes-Haydock variation principle.\cite{rf:foulkes1989prb} The direct optimization of the potential through the increments of the difference between the current and the target densities will be advantageous for constructing the detailed structures of the potential. The long-range nature of the exchange potential was fully realized by including the Slater's local exchange potential in the initial guess for the potential optimizations in the inv-KS as well as in the HF-OEP. 

The inv-KS method implemented with the RSG approach was also applied to an ortho-benzyne radical (C$_6$H$_4$) known as a strongly correlated polyatomic molecule. It was demonstrated that the exchange potentials $\upsilon_{x\sigma}$ for the spin-polarized system can be constructed so that the resultant spin populations assigned to the atomic sites faithfully agree with those obtained by the UHF calculations. Thus it was revealed that the inv-KS approach for the HF electron density can optimize the exchange potential with better accuracies and with a much less effort for the computer implementation as compared with the HF-OEP. 

We make a remark on the extension of the inv-KS method to the densities constructed from the multi-configurational wave functions obtained e.g. by the multi-configuration self-consistent field (MCSCF) method. Provided that the density given gy the MCSCF wave function is $\upsilon$-representable, the local effective potential can also be obtained by the inv-KS. However, it may not be straightforward to evaluate the corresponding total energy because the wave function in the inv-KS is described with a single determinant. On the other hand, it is possible to extend the OEP approach to the MCSCF wave function (MC-OEP) as was performed by Weimer et al in Ref. \cite{rf:weimer2008jcp}. In this sense, the inv-KS approach must be considered in a different context from the OEP method. Actually, the density subspace covered by the wave functions of the MC-OEP approach is the MC-OEP $\upsilon$-rep. density and it includes the subspace composed of the non-interacting $\upsilon$-rep. densities. 

We also make a brief remark on a future prospect. We are now undertaking a study to extend the inv-KS and the HF-OEP to the KS-DFT based on a novel theoretical framework\cite{rf:Takahashi2018,Takahashi2020jpb} that utilizes the energy electron density as a fundamental variable. Provided that the extension to the new theory can be successfully performed, it may serve as a proof of principle for the theory.        

%\newpage

\begin{acknowledgments}
This paper was supported by the Grant-in-Aid for Scientific Research on Innovative Areas (No. 23118701) from the Ministry of Education, Culture, Sports, Science, and Technology (MEXT); the Grant-in-Aid for Challenging Exploratory Research (No. 25620004) from the Japan Society for the Promotion of Science (JSPS); and the Grant-in-Aid for Scientific Research(C) (No. 17K05138 and No. 22K12055) from the Japan Society for the Promotion of Science (JSPS). %This research also used computational resources of the HPCI system provided by Kyoto, Nagoya, and Osaka university through the HPCI System Research Project (Project IDs: hp170046, hp180030, hp180032, hp190011, and hp200016).
\end{acknowledgments}  
% The \nocite command causes all entries in a bibliography to be printed out
% whether or not they are actually referenced in the text. This is appropriate
% for the sample file to show the different styles of references, but authors
% most likely will not want to use it.
%\nocite{*}

%\bibliography{apssamp}% Produces the bibliography via BibTeX.

%apsrev4-2.bst 2019-01-14 (MD) hand-edited version of apsrev4-1.bst
%Control: key (0)
%Control: author (8) initials jnrlst
%Control: editor formatted (1) identically to author
%Control: production of article title (0) allowed
%Control: page (0) single
%Control: year (1) truncated
%Control: production of eprint (0) enabled
\providecommand{\noopsort}[1]{}\providecommand{\singleletter}[1]{#1}%

\end{document}